# Data Challenges for Next-generation Radio Telescopes

Ray P. Norris,
CSIRO Astronomy & Space Science
Sydney, Australia
Ray.Norris@csiro.au

*Abstract*— Radio-astronomy is about to embark on a new way of doing science. The revolution that is about to take place is not due to the enormous sensitivity of the Square Kilometre Array, which is still a decade away, but due to its pathfinders, which are pioneering new ways of doing radio-astronomy. These new ways include multi-pixel phased-array feeds, the goal of producing science-ready images from a real-time pipeline processor, and from the vast amounts of survey data that will be available in the public domain soon after observing. Here I review the data challenges that need to be addressed if we are to reap all the science that potentially resides in SKA Pathfinder data. Some challenges are obvious, such as petabytes of data storage, and some are less obvious, such as the techniques we have yet to develop to perform cross-identifications on millions of galaxies.

*Keywords: astronomy; data; radio-astronomy*

## I. INTRODUCTION

The Square Kilometre Array (SKA) is a $2 billion internationally-funded radio-telescope to be built in either Australia or South Africa [1]. $140m has already been committed to its development, and it is hoped to start construction in 2012-15, with completion in 2022. It will be about 100 times more sensitive than any existing radio telescope, and it will answer fundamental questions about the Universe [2]. It is likely to consist of between 1000 and 1500 15-meter dishes in a central 5km area, surrounded by an equal number of dishes in a region stretching up to thousands of km. All these dishes will be connected to a massive real-time data processor by an optical fibre network.

From an initial short list of 6 possible host countries, the choice has been narrowed down to two (Australia or South Africa) with the final decision between them to be made in 2012. The primary site selection criterion is radio-quietness. Current radio-telescopes throughout the world are increasingly swamped by interference from TV transmitters, mobile phones, Wi-Fi, and all the other wireless accoutrements of modern life. Very few places in the world are free of the radio interference generated by them, and when combined with the requirement of access to modern infrastructure in a politically stable country, the choice of SKA site is reduced to two. For example, the Australian candidate SKA site, in Murchison shire in Western Australia, has a population of less than 160 people in a region the size of the Netherlands, and yet is sufficiently close to civilisation to be able to lay a high-bandwidth fibre to the site. Just as very few next-generation large optical telescopes are being built anywhere other than Atacama (Chile) or Hawaii, it is unlikely that any large next-generation radio telescopes will now be built anywhere other than Australia or South Africa.

## II. THE SKA PATHFINDERS

This paper is not about the SKA, but about its Pathfinders. The SKA Pathfinders are new telescopes, or upgrades of existing telescopes, that are being built both to test and develop aspects of potential SKA technology and to develop SKA science. They include the Australian SKA Pathfinder (ASKAP), the South African MeerKAT telescope, the LOFAR array in the Netherlands, the Apertif upgrade to the existing Westerbork telescope in The Netherlands, the Allen Telescope Array (ATA) in the USA, the Murchison Wide field Array (MWA) in Australia, the Extended Very Large Array (EVLA) in the USA, and eMERLIN in the UK. Here I focus mainly on ASKAP, which is being built on the Australian candidate SKA site in Western Australia, but most of the comments apply equally to the other pathfinders. Each of them shares the characteristics that in addition to developing SKA science or technology, each is a major telescope in its own right, likely to generate significant new astronomical discoveries [3].

For example, ASKAP, shown in Fig. 1, consists of 36 12-metre antennas spread over a region 8 km in diameter. It is fully funded, and construction has started, with completion expected in 2012. The array of antennas is no larger than many existing radio-telescopes, but ASKAP's feed array is revolutionary, with a phased-array feed of about 100 pixels replacing the single-pixel feeds that are almost universal in current-generation radio-telescopes. As a result, it will have a thirty square degree field of view, so can survey the sky some thirty times faster than existing radio-telescopes, enabling surveys that could not currently be contemplated.

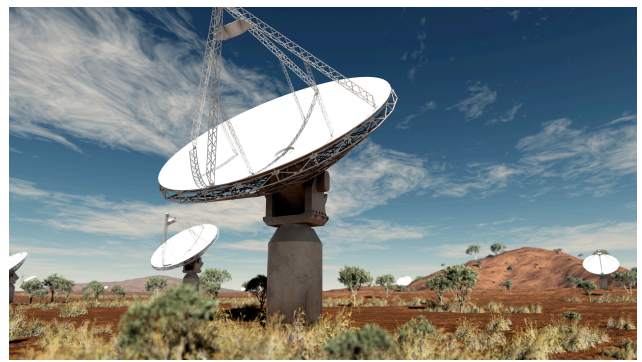

Figure 1. An artist's impression of the Australian SKA pathfinder telescope, due to be completed in 2012. Shown are a few of the 36 12-metre dishes, each equipped with a 100-pixel Phased Array Feed. Image courtesy of CSIRO.

### III. SURVEY OBSERVATIONS

Traditionally, an astronomer wishing to understand a new type of astronomical object would apply for observing time on many different telescopes operating at different wavelengths. To an increasing extent, astronomers can now access data from large-scale surveys at a depth and resolution comparable to that obtained from a conventional observation on a large telescope. This is changing the way that we do astronomy [4].

In twenty years time, it is likely that all the sky will have been imaged deeply and catalogued by major survey projects, and so an astronomer will be able to access deep multi-wavelength data on any object being studied. Even more importantly, the astronomer will be able to access large numbers of such objects. As a result, astronomical discoveries will not only come from new telescopes observing new regions of observational phase space, but also from astronomers who mine the databases in innovative ways, and stack images to unprecedented sensitivities.

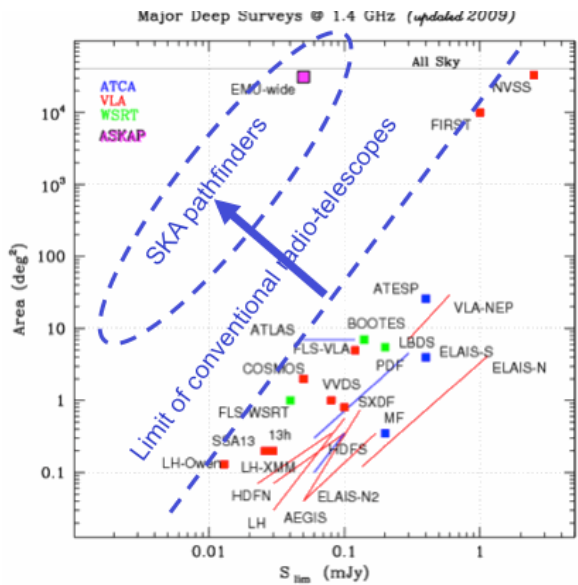

Figure 2. The observational phase space of 20cm radio continuum surveys, taken from [4], and originally generated by Isabella Prandoni. The most sensitive surveys are on the bottom left, and the largest area surveys on the top right. The dashed diagonal line shows the limiting envelope of existing surveys, which are largely limited by available telescope time. Next-generation SKA pathfinders such as the Australian SKA Pathfinder (ASKAP) are able to survey the unexplored region of phase space to the top left by using innovative approaches such as focal plane arrays.

Fig. 2 shows current 20-cm continuum radio surveys, ranging from the wide but shallow NRAO VLA Sky Survey (NVSS) [5]; in the top right to the deep but narrow Lockman Hole observation [6] in the lower left. The region occupied by these surveys has been roughly uniformly populated with new discoveries, and even the most recent large survey (ATLAS) [7,8] has been able to discover a new class of radio galaxy: the Infrared-Faint Radio Sources [9,10]. We have no reason to doubt that the remaining half of this diagram is equally populated with new discoveries, and yet we are currently prevented from venturing into that unexplored white space by the availability of telescope time. The enormously increased speed of the SKA Pathfinders opens up this area, enabling us to survey where no telescope has surveyed before, doubtless discovering new phenomena, and perhaps solving some of astronomy's most pressing questions.

Each of the SKA Pathfinder telescopes is engaged in a number of surveys, and in each case the suite of surveys includes a major continuum survey, which will explore the white space to the left of Fig. 2.

### IV. THE EMU SURVEY

To illustrate how such large survey projects work, here I consider one such survey, the Evolutionary Map of the Universe (EMU)[11]. EMU is a wide-field radio continuum survey planned for ASKAP, and is expected to take about 1.5 years of observing time. The primary goal of EMU is to make a deep (~10μJy rms) radio continuum survey of the entire Southern Sky, extending as far North as declination +30°. EMU is expected to detect and catalogue about 70 million galaxies, including typical star forming galaxies up to a redshift of 1, powerful starbursts to even greater redshifts, Active Galactic Nuclei (AGN) to the edge of the Universe, and will undoubtedly discover new classes of object. Surveys at this depth extend beyond the traditional domains of radio astronomy, dominated by radio-loud galaxies and quasars, into the regime of star-forming galaxies previously dominated by ultraviolet, optical, and infrared surveys.

EMU may be compared with the NRAO VLA Sky Survey [5] but will be 45 times more sensitive than NVSS and will have an angular resolution (10 arcsec) five times higher, whilst covering a similar area (75% of the sky). All radio data from the survey will be placed in the public domain as soon as the data quality has been checked.

A further difference from previous large radio surveys is that an integral part of the EMU project will be to cross-identify the EMU radio sources with sources in major surveys at other wavelengths, and produce public-domain VO-accessible catalogues as "added-value" data products. This is facilitated by the growth in the number of large southern hemisphere telescopes and associated planned major surveys spanning all wavelengths, discussed below.

Broadly, the key science goals for EMU are:
- To trace the evolution of star-forming galaxies from z=2 to the present day, using a wavelength unbiased by dust or molecular emission.
- To trace the evolution of massive black holes throughout the history of the Universe, and understand their relationship to star-formation.
- To use the distribution of radio sources to explore the large-scale structure and cosmological parameters of the Universe.
- To explore an uncharted region of observational parameter space, with a high likelihood of finding new classes of object.
- To determine how radio sources populate dark matter halos, as a step towards understanding the underlying astrophysics of clusters and halos.
- To create the most sensitive wide-field atlas of Galactic continuum emission yet made in the Southern Hemisphere, addressing areas such as star formation, supernovae, and Galactic structure.

## V. DATA CHALLENGES

Not only are large surveys such as EMU changing the observing mode, but the way that we process the data is changing too. Until recently, astronomers would observe at the telescope, then take their data home on disks and spend weeks or months processing it. However, next-generation large surveys will generate data volumes that are too large to transport, and so all processing must be done in situ. To keep up with the data flow from the telescope, data processing must be done in near-real-time, which requires that it must be automated. Almost all next-generation telescopes now being designed offer automated pipeline reduction of data.

TABLE I. THE DATA CHALLENGES OF SKA PATHFINDERS.
Figures shown are budgeted for ASKAP and will be delivered by 2012. Requirements for the SKA itself are about 100 times greater.

| Challenge | Specifications |
|---|---|
| Large bandwidth from telescope to processor | ~10 Tb/s from antennas to correlator (< 6 km)<br>40 Gb/s from correlator to processor (~ 600 km) |
| Large processing power | 750 Tflop/s expected/budgeted<br>1 Pflop desired |
| Power consumption of processors | 1 MW at site,<br>10 MW for processor<br>power bill ~$3M p.a. |
| Pipeline processing essential | including data validation, source extraction, cross-identification, etc |
| Storage and curation of data | 70 PB/yr if all products are kept<br>5 PB/yr with current funding<br>8 h to write 12 h of data to disk at 10GB/s |
| Retrieval of data by users | all data in public domain<br>accessed using VO tools & services |
| Data-intensive research | data mining, stacking, cross-correlation, etc. |

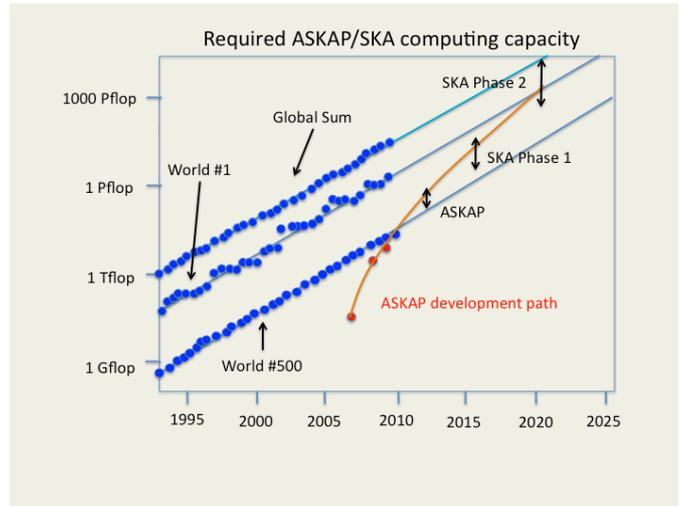

Figure 3: Available computing power over time compared with ASKAP and SKA data needs, adapted from a figure by Tim Cornwell. The three lines show the world's fastest computer, the world's 500th most powerful computer, and the sum of the top 500 computers.

Table 1 shows the data challenges [12] for ASKAP. Other SKA Pathfinders are likely to have similar data challenges, and SKA itself will have data rates perhaps 100 times greater.

Meeting these data challenges is barely possible even if Moore's Law continues to hold. Fig. 3 shows a plot of processor power against time. The SKA in 2022 will require the most powerful computer in the world to meet its data challenges.

Another trend is that data are increasingly being placed in the public domain. Whilst proprietary periods are necessary in some cases, much astronomical discovery occurs after the data are released to other groups, who are able to add value to the data by combining it with data, models, or ideas that may not have been accessible to the survey astronomers.

## VI. DATA ANALYSIS AND CROSS-IDENTIFICATION

Public-domain multi-wavelength surveys such as EMU enable an astronomer to study large numbers of astronomical objects at many wavelengths. As a result, astronomical discoveries are increasingly coming from astronomers who mine the databases in innovative ways, and stack images to unprecedented sensitivities. The insight and creativity that used to be focused on the best observing proposals is increasingly focusing on innovative ways of extracting data from the databases. In a sense, we are still observing, but our observations are increasingly of a virtual sky rather than the real sky.

Just as data volumes become unmanageable for the computing resources of an individual astronomer, the volumes of processed information become unmanageable for the astronomer's brain. Comparison of radio and optical images has traditionally been done by eye, but this will become impossible in all but a few rare cases. For example, EMU is expected to discover and catalogue about 70 million galaxies at radio wavelengths. The science will only flow from these catalogues when we cross-identify them with optical/infrared objects from surveys such as WISE [13], VISTA [14], Sloan [15], and SkyMapper [16]. We expect that 70% of the cross-identifications will be performed automatically with a simple algorithm which compares catalogues for simple isolated objects, and assigns probabilities to cross-matches based on their separation, their properties (such as brightness, colour, or polarisation), and our prior knowledge of astrophysical objects. For extended or multiple objects, the process will be much harder, and at present we do not have algorithms that can handle these hard questions, and the tens of millions of cross-identifications preclude human cross-identification by the astronomer.

Projects such as Galaxy Zoo [17] already tackle a similar problem, that of galaxy classification from the Sloan survey, by employing enthusiastic citizens from around the world who would like to contribute to the scientific endeavour. Users are led through a simple decision tree to make a complex classification of a galaxy. Even an untrained human brain is far superior to the best algorithm currently available, and Galaxy Zoo success rates are high, with real science being generated. The EMU team are collaborating with the Galaxy Zoo team to apply their technology and experience to the problem of cross-matching radio sources.

## VII. DATA-INTENSIVE RESEARCH

These large volumes of survey data will enable types of data analysis that are currently beyond the reach of conventional astronomy. Examples include

- **Data mining** e.g. searching for correlations between radio properties and other multi-wavelength properties.
- **Stacking** i.e. adding together results from many different instances of one class of object, to measure the properties of the class as a whole. Since millions of objects can be stacked, this can reach sensitivities far in excess of the original data (e.g. [18]).
- **Cross-correlation** i.e. correlating the positions of radio objects with object or environments found in a different survey. For example, the evolution of dark energy can in principle be measured by cross-correlating radio source positions with the cosmic microwave background, using the Integrated Sachs-Wolf effect.

Not only are these tasks compute intensive (perhaps requiring as much processing power as the original data processing) but they also need high bandwidth to the various data sets, which therefore either need to be geographically co-located, or else have very high bandwidth links (e.g. 50 GB/s sustainable) plus hundreds of TB of disk. It is not yet clear whether these resources will be available. If not, we will have lost a great opportunity to extract the science from ASKAP data.

## VIII. CONCLUSION

Next generation radio telescopes offer us a profound change in the way that we do astronomy, with the promise of major discoveries by exploring fresh areas of observational phase space and applying new data-intensive techniques to extract the science from the data. But they will also require a level of computing power that place them amongst the most computer-hungry projects on Earth and also assume that our current approaches are all scalable. These are enormous challenges, and no doubt we will encounter some surprises along the way. However, if we can get it right, the scientific rewards are likely to be immense.

[1] Dewdney, P., et al., 2009, "The Sqaure Kilometre Array", Proc. IEEE, Vol. 97, pp. 1482-1496.

[2] Carilli, C., & Rawlings, S., 2004, "Science with the Square Kilometre Array", New Astronomy Reviews, Vol. 48, Elsevier.

[3] Johnston, S.; et al. 2008, "Science with ASKAP", Experimental Astronomy, Vol. 22., pp 151-273.

[4] Norris, R. P., 2010, "Next-Generation Astronomy", in "Accelerating The Rate Of Astronomical Discovery", edited by Ray P. Norris & Clive L. N. Ruggle, 2010, PoS(SpS5), http://arxiv.org/abs/1009.6027

[5] Condon, J., et al. 1998, "The NRAO VLA Sky Survey", Astron. J, Vol. 115, 1693-1716.

[6] Owen, F.N, Morison, G.E., 2008, "The Deep Swire Field", Astron. Journal, Volume 136, pp. 1889-1900

[7] Norris, R.P., et al., 2006, "Deep ATLAS radio observations of the CDFS-SWIRE field", 2006, Astron. J, 132, 2409-2423.

[8] Middelberg, E., et al., 2008, "Deep ATLAS Radio Observations of the ELAIS-S1 field", Astron. J, 135.1276-1290.

[9] Middelberg, E., et al., 2010, "The Radio Properties of Infrared-Faint Radio Sources", Astron. Astrophys, in press.

[10] Norris, R.P., et al., 2010, " Deep Spitzer Observations of Infrared-Faint Radio Sources: High-redshift Radio-Loud AGN?", Astrophys. J, in press.

[11] Norris R.P., et al., 2009, "Evolutionary Map of the Universe", in "Proceedings of Panoramic Radio Astronomy". Ed George Heald. PoS(PRA2009)033, arXiv:0909.3666

[12] Cornwell, T.J., 2010, pers. communication.

[13] Liu, F., et al. 2008,Proceedings of the SPIE, Vol. 7017, pp. 70170M-70170M-12

[14] Sutherland, W. 2009, "VISTA Public Surveys and VLT followup" in "Science with the VLT in the ELT Era", Astrophysics and Space Science Proceedings, p. 171-175.

[15] Abazajian, K.N., et al. 2009, " The Seventh Data Release of the Sloan Digital Sky Survey", Astrophysical Journal Supplement, Vol. 182, pp. 543-558

[16] Keller, S. C., et al. 2007, "The SkyMapper Telescope and The Southern Sky Survey", Publications of the Astronomical Society of Australia, Vol. 24, pp. 1-12.

[17] Lintott, C. J., et al. 2008, " Galaxy Zoo: morphologies derived from visual inspection of galaxies from the Sloan Digital Sky Survey", Monthly Notices of the Royal Astronomical Society, Volume 389, pp. 1179-1189

[18] Boyle, B.J., et al., 2007, "Extending the Infrared-Radio correlation", Mon. Not. R. Astr. Soc., Vol. 376, pp1182-1188